\documentclass[final]{interact}

\usepackage{epstopdf}
\usepackage[caption=false]{subfig}

\usepackage{bbding}
\usepackage{pifont}
\usepackage{amssymb}

\usepackage[natbibapa,nodoi]{apacite}
\theoremstyle{plain}

\theoremstyle{definition}

\theoremstyle{remark}

\usepackage{pdfpages}

\usepackage{nameref}
\usepackage{varioref}
\usepackage{hyperref}
\usepackage{cleveref}

\usepackage{threeparttable,booktabs}

\usepackage{tikz}

\newcommand\copyrighttext{%
	\footnotesize This is an Accepted Manuscript of an article published by Taylor \& Francis in \textbf{Interactive Learning Environments} on 9 Nov 2021, available online: \url{https//doi.org/10.1080/10494820.2021.1999985}.”}
\newcommand\copyrightnotice{%
	\begin{tikzpicture}[remember picture,overlay]
		\node[anchor=south,yshift=10pt] at (current page.south) {\fbox{\parbox{\dimexpr\textwidth-\fboxsep-\fboxrule\relax}{\copyrighttext}}};
	\end{tikzpicture}%
}

\begin{document}

\articletype{Interactive Learning Environments}

\title{An Adaptive 3D Virtual Learning Environment for Training Software Developers in Scrum}

\author{
\name{Ezequiel Scott\textsuperscript{a}\thanks{CONTACT Ezequiel Scott. Email: ezequiel.scott@ut.ee} and Marcelo Campo\textsuperscript{b,c}}
\affil{\textsuperscript{a} University of Tartu, Institute of Computer Science, Tartu, Estonia;\\ \textsuperscript{b} NICE, Facultad de Ciencias Exactas, UNICEN National University, Buenos Aires, Tandil, Argentina; \\
\textsuperscript{c} CONICET, Consejo Nacional de Investigaciones Cientificas y Tecnicas, Buenos Aires, Argentina}
}

\maketitle

\begin{abstract}
New methods for software development have been introduced in the last years. Scrum, in particular, is one of the most used frameworks for software development because of its potential improvements in productivity, quality, and client satisfaction. Academia has also focused on teaching Scrum practices to prepare students to face common software engineering challenges and facilitate their insertion in professional contexts. Furthermore, advances in learning technologies currently offer many virtual learning environments to enhance learning in many ways. Their capability to consider the individual learner preferences has led a shift to more personalised training approaches, requiring that the environments adapt themselves to the learner. In this context, we propose an adaptive approach for training Scrum software developers to improve their understanding of the Scrum framework. Our approach to train developers uses an adaptive virtual learning environment based on Felder's learning style theory. Although still preliminary, our findings show that students who used the environment and received instruction matching their preferences obtained sightly higher learning gains than students who received a different instruction than the one they prefer. We also noticed less variability in the learning gains of students who received instruction matching their preferences. The relevance of this work goes beyond their impact learning gains due to the virtual and collaborative features of the environment, describing how adaptive virtual learning environments can be used in the domain of Software Engineering.
\end{abstract}


\copyrightnotice

\section{Introduction}

Teaching agile practices is at the cutting-edge of Software Engineering education since agile methodologies are widely used in the industry. As university-level teaching aims at preparing students for both facing common software engineering challenges and facilitating their insertion in professional contexts, the software engineering curricula should respond by providing students with knowledge and meaningful experiences on how to use Scrum in practice. Therefore, there is an increasing need to introduce Scrum into training courses, and many educators are faced with how to incorporate Scrum into their software engineering courses.

Over the last years, researchers have focused on teaching Scrum practices and have proposed many approaches. These go beyond the traditional lectures in which all the students listen to the teacher and eventually take notes because, as a general consensus, educators claim that teaching Scrum should include some practical experience to strengthen comprehension and achieve deep learning \citep{Mahnic2015}. For this reason, the use of Scrum in capstone projects is the most adopted strategy worldwide. However, capstone projects may still preserve characteristics of one-size-fits-all education, where the same teaching strategies are considered suitable for all kinds of learners.

For this reason, it is crucial to provide students with personalised learning scenarios that might offer more meaningful learning experiences. The personalisation of the learning process is aligned with learner-centred principles, in which professors should consider students' learning preferences, their strengths and weaknesses \citep{McCombs1997}. In this context, learning styles are not only useful tools to detect how learners perceive, interact with, and respond to the learning environment but also valuable indicators for the cognitive, affective, and psychological students' behaviours \citep{Keefe1988}. Moreover, previous studies have shown that students' learning experiences related to Scrum can be enhanced along with their educational outcomes when learning styles are considered \citep{scott2016towards,Scott2014}.

This paper proposes an adaptive 3D virtual learning environment in which Scrum topics are tailored to the developer's learning style. We model the developer's profile using the Felder-Silverman learning style model \citep{Felder1988}. The profile is used by an adaptation mechanism responsible for adapting the virtual environment interface and providing learning content to the developer conforming to their learning style. The adaptation mechanism is included in a virtual environment that supports software development based on Scrum, namely Virtual Scrum \citep{Rodriguez2013}, which has shown favourable results in previous works and in the context of teaching Scrum. We conducted a controlled experiment with students, and our results show that students who used the environment and received instruction matching their preferences obtained slightly higher learning gains than students who received a different instruction than the one they prefer. These results are promising since they support the use of adaptive virtual learning environments in the domain of Software Engineering.

\section{Related Work} \label{sec:related-work}

\subsection{Virtual Worlds in Education}

Virtual Learning Environments (VLEs) have the specific purpose of enabling teaching and learning by providing valuable features for students and teachers \citep{Aldrich2009}. Virtual Environments can also be considered part of a broader category: Virtual Worlds, as defined by \cite{nevelsteen2018virtual}. Virtual World is a more popular term that groups several interactive environments with different characteristics and properties. Virtual Worlds have been attractive to researchers in the educational field due to their potential benefits from a technological perspective.

Virtual worlds present unique characteristics that allow educators to achieve several ongoing challenges. Many recent studies \citep{lau2015use, Park2003,kavanagh2017systematic, reisouglu20173d, Warburton2009, Duncan2012} have shown that Virtual Worlds can contribute to engagement and interactivity, collaboration, experimentation, formation of group identity, and idea generation. Virtual worlds aim to foster students to participate in simulated activities and collaborate in distributed settings, similar to how they do face-to-face activities.

These benefits are partially due to the high fidelity to the reality that virtual worlds offer. For this reason, there is a trend towards building complex virtual environments in which both pedagogical and gaming elements take relevance. Numerous technical requirements challenge the development of virtual worlds with these characteristics; nevertheless, several technological solutions and platforms make creating Virtual Worlds easier.

Vircadia™\footnote{https://github.com/vircadia/vircadia} is a 3D open-source multiplayer software project that allows for creating social and educational environments. Vircadia consists of many projects and hundreds of contributors. Although the platform is relatively recent, researchers are paying attention to its capabilities for Virtual Reality \citep{kwon2020creative}.
VoRtex \citep{jovanovic2019vortex} is another platform for creating Virtual Worlds. One of the main elements of the platform is Vortex School\footnote{https://github.com/Aca1990/VoRtex-School}, which adapts the concept of MicroLessons. MicroLessons include challenges like problem-solving puzzles for different use cases such as learning a new programming algorithm. This platform is free and open-source. Among other open-source initiatives for creating Virtual Worlds, we can mention Tivoli Cloud VR\footnote{https://github.com/tivolicloud/} and Sansar\footnote{https://www.sansar.com/}. 
Popular games have also introduced creative modes to enable the use of a customized version of the game in the classroom. For instance, Fortnite Creative\footnote{https://www.unrealengine.com/en-US/onlinelearning-courses/teaching-with-fortnite-creative} allows teachers to build a lesson plan to teach content-specific learning outcomes.  
Second Life\footnote{https://secondlife.com/} is probably the popular and mature platform for creating Virtual Worlds \citep{reisouglu20173d, Warburton2009}. Second Life has many users that use the platform for many purposes, ranging from entertainment to education. 

Virtual Worlds for education also present several issues and limitations that current studies have reported. For instance, \cite{kavanagh2017systematic} report on problems related to cost, training, and software and hardware usability. The authors also reported that students found the implementations to be insufficiently realistic. \cite{girvan2012ethical} point out that Virtual worlds provide the researchers with new ethical issues to consider, such as obtaining informed consent, privacy protection and identity. In a meta-review \citep{reisouglu20173d}, the authors remark that virtual worlds should focus on developing high-order cognitive skills. Moreover, a better evaluation of these environments is needed to reveal students' achievements more clearly. These findings are in line with other studies in the field \citep{Scott2016}. Despite these issues, when virtual worlds are carefully designed and evaluated, their outcomes can be encouraging, which make them still prominent as a learning technology.

\subsection{Approaches for Teaching Scrum}

The Scrum framework is simple to understand yet difficult to master \citep{Schwaber2002}. For this reason, many approaches have arisen to address this issue. There are mainly three directions that go beyond the traditional lectures on Scrum: teaching Scrum through the use of games, through the experience of capstone projects, and using the aid of virtual environments, sometimes including games. When Scrum is taught using games, learners follow the well-defined rules of the game in a partial or complete simulation of the software development process. These games are generally suitable for learners as their opinions about the learning process have been satisfactory. There are currently several examples of these learning approaches, such as the work of \citep{Wangenheim2013, Paasivaara2014, Ramingwong2015, Krivitsky2011, Fernandes2010, guzman2019teaching, baumann2020teaching}.

Another direction proposes courses that take advantage of capstone projects, in which professional working environments are simulated as much as possible. These courses require learners to work in teams to develop a software project, usually within the scope of a software engineering course. Capstone projects present many advantages, such as helping students to develop transferable skills that are better acquired within a practical project rather than through formal lectures. Several articles report the use of capstone projects for teaching Scrum \citep{Mahnic2012, Mahnic2015, Devedzic2011, Rico2009, Tan2008}.

The third direction comprises different approaches in which software tools are crucial to support Scrum practices in educational and professional contexts \citep{WisnuWirawan2014, Yu2014}. Several 2D tools can be used to support the teaching of Scrum, for instance \citep{lee2016scrum, begosso2019simscrumf}. However, 2D tools usually fail to represent the physical development environment to carry out a realistic simulation of the Scrum process. Virtual worlds arise as an adequate solution to address this problem because there is a lack of physical space and resources that prevent creating Scrum team rooms in universities. Moreover, recent studies have shown that Virtual Learning Environments rapidly become an integral part of teaching and learning in higher education \citep{Callaghan2009}. Video games and virtual worlds are moving into the mainstream as traditional media industries struggle to keep with up digital natives and their desire for information, technology and connectivity. Researchers in the field are convinced that these technologies are potentially profitable to education \citep{Anderson2008, Rodriguez2012c}.

In the light of the above, 3D virtual environments allow developers to know about tasks performed by their peers and even hold meetings regardless of their physical location. Also, these environments can be used to provide a physical topology of both a software project and a process. This characteristic allows for faster access to information than a 2D tool plus video conference system to carry out meetings within a distributed team \citep{Whitehead2007}. This benefit may be attributed to the possibility of integrating isolated tools in a single development environment, leading to a context in which it is viable to share software artefacts. Therefore, using a virtual world has many advantages in comparison with 2D tools. For instance, students use avatars to communicate through gestures \citep{Herbsleb2007}, and they also can manipulate elements in the modelled room going through a 3D experience of Scrum. Additionally, the sense of immersion and the simulated team room, equipped with the Scrum artefacts and rules, is crucial to engaging users in a hands-on experience of using Scrum, participate in collaborative work, and improve their comprehension of this agile method.

Several studies have used virtual environments for teaching Scrum. For instance, the recent work of \cite{radhakrishnan2020teaching} proposes creating a collaborative virtual simulation of Scrum named VRScrum. SimScrum \citep{begosso2019simscrumf} is another work in progress that proposes to use gamification principles to create a simulator for teaching Scrum in Software Engineering courses. There are also more mature projects with promising results. For instance, ScrumVR \cite{caserman2020become} is an educational game to teach Scrum through a game scenario where players have to participate in activities. The game uses virtual reality technologies, and experiments show that it is helpful to teach the foundations of Scrum. \cite{lee2016scrum} proposes Scrum-X, another simulation game to teach Scrum. During the game, players can plan, execute and manage a software development project, and they will be able to experience the entire Scrum process. The author conducted a pilot study with promising results.

Virtual Scrum \citep{Rodriguez2013} uses a virtual world to simulate a work environment handling 3D displays of the Scrum artefacts. Virtual Scrum also supports a task board for planning and tracking user stories; a daily meeting artefact for solving problems, adapting rapidly to changes and removing impediments; and a burn-down chart for reflecting on the past Sprint and making continuous process improvements during retrospective meetings. This way, Virtual Scrum uses a virtual world to harness the 3D interfaces for training students in their performance in a simulated Scrum environment. 

Table \ref{tab:features} summarizes the recent virtual environments for teaching Scrum and compares them. We included JIRA in the table because of its popularity and extensive use for project management, although the main purpose of JIRA is not educational; thus, the tool serves as a baseline for feature comparison. Table \ref{tab:features} shows that student profiling and simulated scenarios are the less explored features when teaching Scrum using virtual environments.

In summary, there have been several research directions for teaching Scrum in academic and professional contexts. Although many approaches have shown improvements from many perspectives, they have neglected personalising issues. A recent literature review indicates that adaptive 3D virtual learning environments \citep{Scott2016} have shown to be suitable software tools to support personalised learning, providing learners with unique learning experiences due to their 3D features. Adaptive 3D VLEs have been used in many domains, but to the best of our knowledge, a specific adaptive environment still has not been developed for training developers in Scrum. In this paper, we introduce Adaptive Virtual Scrum, an extension of Virtual Scrum \citep{Rodriguez2013}, which includes an adaptation mechanism to support the personalised learning of Scrum. 

\begin{table}[]
\caption{Feature table comparison.}
\footnotesize{
\label{tab:features}
\begin{threeparttable}
\begin{tabular}{@{}p{4cm}p{.5cm}p{1.2cm}p{1.4cm}p{1.3cm}p{1.3cm}p{1cm}p{1.3cm}@{}}
\toprule
Feature 	&	 Jira\tnote{1} 	&	 Scrum-X\tnote{2} 	&	 SimScrumF\tnote{3} 	&	 VRScrum\tnote{4} 	&	 ScrumVR\tnote{5} 	&	 Virtual Scrum\tnote{6} 	&	 Adaptive Virtual Scrum \\ \midrule
3D environment 	&	  	&	 \Checkmark 	&	 \Checkmark 	&	 \Checkmark 	&	 \Checkmark 	&	 \Checkmark 	&	 \Checkmark \\
Use of avatars 	&	  	&	  	&	  	&	 \Checkmark 	&	 \Checkmark 	&	 \Checkmark 	&	 \Checkmark \\
Embedded chat 	&	  	&	  	&	  	&	  	&	  	&	 \Checkmark 	&	 \Checkmark \\
Role support 	&	  	&	 \Checkmark 	&	 \Checkmark 	&	 \Checkmark 	&	 \Checkmark 	&	 \Checkmark	&	 \Checkmark\\
Calendar view 	&	 \Checkmark 	&	  	&	  	&	  	&	  	&	 \Checkmark 	&	  \\
Product backlog 	&	 \Checkmark	&	 \Checkmark	&	 \Checkmark	&	 \Checkmark	&	 \Checkmark	&	 \Checkmark 	&	 \Checkmark \\
User stories 	&	 \Checkmark 	&	 \Checkmark 	&	 \Checkmark	&	 \Checkmark	&	 \Checkmark	&	 \Checkmark	&	 \Checkmark\\
Planning Poker View 	&	  	&	  	&	  	&	  	&	  	&	 \Checkmark 	&	 \Checkmark\\
Daily Meeting View 	&	 \Checkmark 	&	  	&	  	&	  	&	 \Checkmark 	&	 \Checkmark	&  \Checkmark \\
Kanban board & \Checkmark & \Checkmark & \Checkmark & \Checkmark  & & \Checkmark	&	 \Checkmark \\
Sprint review/retrospective 	&	  	&	  	&	  	&	  	&	 \Checkmark	&	 \Checkmark	&	 \Checkmark\\
Burndown chart 	&	 \Checkmark	&	 \Checkmark	&	  	&	  	&	  	&	 \Checkmark	&	 \Checkmark\\
Learning activities 	&	  	&	 \Checkmark 	&	  	&	  	&	 \Checkmark	&	  	&	 \Checkmark\\
Adaptation mechanism 	&	  	&	  	&	  	&	  	&	  	&	  	&	 \Checkmark\\
Multiple projects/teams 	&	 \Checkmark 	&	  	&	 \Checkmark 	&	  	&	 \Checkmark	&	 \Checkmark 	&	 \Checkmark\\
Experience points 	&	  	&	  	&	 \Checkmark 	&	  	&	  	&	  	&	  \\
Student profile 	&	  	&	  	&	 \Checkmark 	&	  	&	  	&	  	&	 \Checkmark \\
Simulated scenarios 	&	  	&	  	&	 \Checkmark 	&	  	&	  	&	  	&	  \\ \bottomrule
\end{tabular}%
\begin{tablenotes}
\item[1] Atlassian website -- \url{https://www.atlassian.com/software/jira}
\item[2] \cite{lee2016scrum} 
\item[3] \cite{begosso2019simscrumf}
\item[4] \cite{radhakrishnan2020teaching}
\item[5] \cite{caserman2020become}
\item[6] \cite{Rodriguez2013} 
\end{tablenotes}
\end{threeparttable}
}
\end{table}

\section{Overview of the approach }\label{sec:Overview}

The approach involves two actors: a developer and a coach. The main actor is the developer, who interacts with the environment to learn the concepts of Scrum. The coach has the responsibility of supervising the learning process. The coach might be a professional trainer in an industrial setting or a university teacher. 

The approach exploits the benefits of capstone projects since they allow students to put Scrum into practise and simulate the professional working environment as much as possible. This way, the approach covers all the phases of the Scrum framework, from organising User Stories to delivering the product. Similarly to capstone projects, the coach sets the initial requirements for the project that students will develop. These requirements can be given to the students through an interview where they have to elicit the requirements directly from the product owner (i.e., the coach), a textual narrative, or even a set of Epics. 

\begin{figure*}
\centering
\includegraphics[width=1\textwidth]{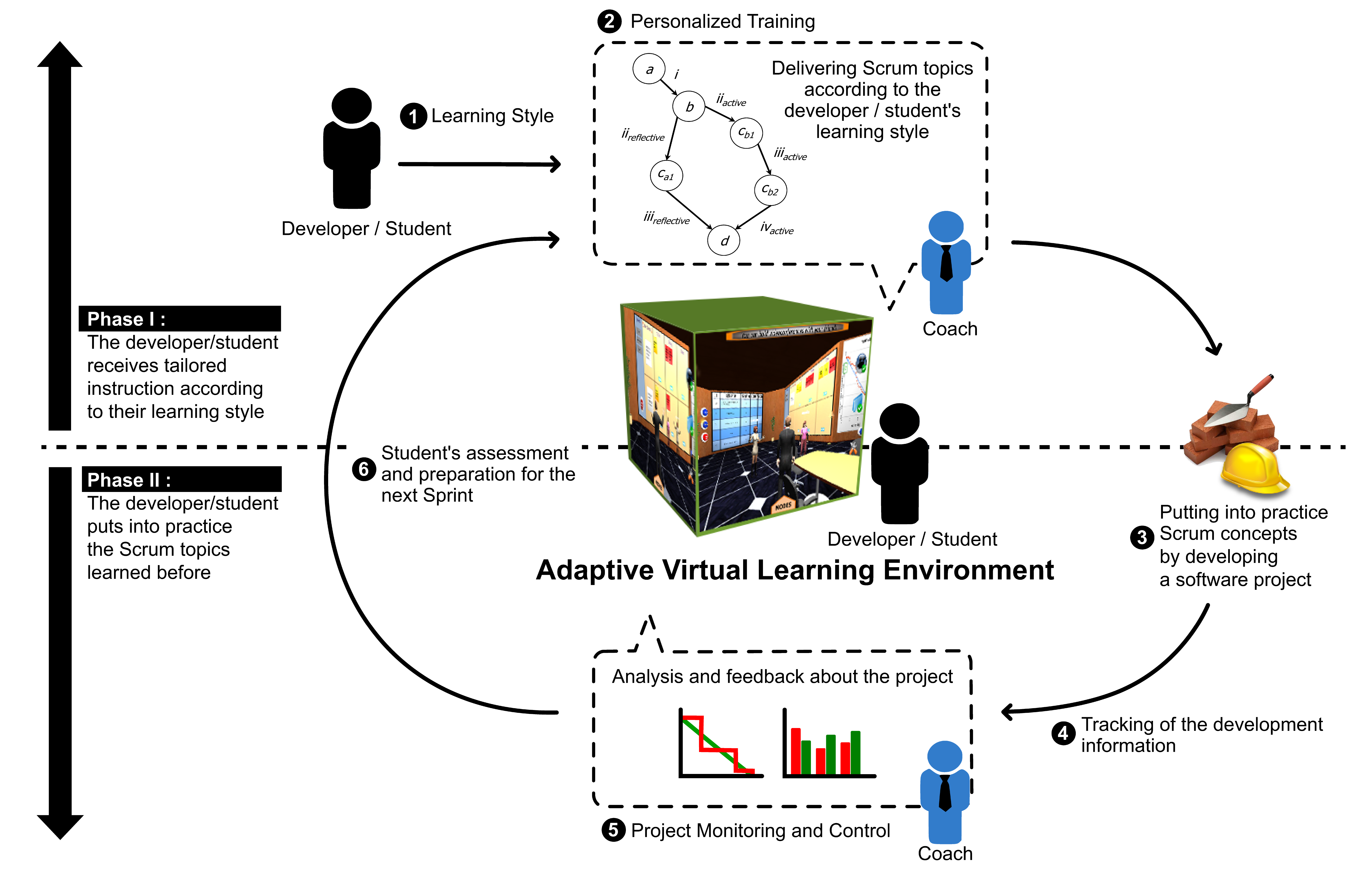}
\caption{The proposed approach.}
\label{fig:Enfoque}
\end{figure*}

Figure \ref{fig:Enfoque} shows an overview of the proposed approach as an iterative process with intermediate steps in which the actors mentioned above are involved. The first step \ding{202} involves knowing the student's learning style according to the Felder-Silverman model \citep{Felder1988}. Currently, these values are obtained from the student's answers to the ILS instrument, but these values may also be obtained through automatic detection of the student's learning style, as shown by several studies \citep{Feldman2014a}. The ILS instrument is a 44-item questionnaire that allows for reliably collecting the student's learning style; the instrument is delivered through an online form connected to the environment. 

The second step \ding{203} is one of the most important since the developer receives content according to their learning style. A component of the environment fires rules for tailoring the user interface according to the previously collected learner model, thereby adapting concepts and learning activities to the student's learning style. Since our environment does not pretend to be an intelligent tutoring system that replaces the role of the teacher, we stress the importance of the role of the coach. The coach could provide the student with valuable feedback and answer additional questions that arise from the explanations of the topics. The environment also helps the coach deliver the Scrum topics because they are already organised and planned with learning activities. 

The third step \ding{204} allows the developer to put the Scrum topics into practice. Working as part of a Scrum Team, the developer participates by following the Scrum topics learned before. The team start with the Sprint planning meeting and end with the Sprint Review and Sprint retrospective meeting. During the Sprint, team members meet regularly at Daily Scrum meetings to inform each other of their current activities and possible impediments. 

The fourth step \ding{205} involves tracking information about how the developers address the Scrum practices since they can record all the information into the artefacts provided by the environment. For example, the environment provides a virtual Product Backlog in which the student can record their specification of the requirements in the form of User Stories and their tasks and estimates. This tracking is done similarly to when developers interact with project management tools such as JIRA\footnote{JIRA Website - https://www.atlassian.com/software/jira}. 

The fifth step \ding{206} takes part at the end of the Sprint, when developers, as a team, have to deliver the product increment to the product owner (i.e., the coach) by performing a Sprint Review meeting. Here, the coach analyses the tracking of the project by interpreting the software development metrics collected in step four; then, she/he makes suggestions and give feedback to the student. Finally, in the sixth step \ding{207}, the developer and the team are ready to advance to the next Sprint, repeating the second step when necessary, yet carrying out steps 3 to 6 until the end of the project. 

\subsection{An illustrative example}

We introduce a simple example to illustrate the idea of the proposed approach. In this example, there are two developers, John and Mary. These developers are part of a Software Engineering Course, and they do not have any prior knowledge about Scrum. During the first step \ding{202}, the developers answer the online version of the ILS questionnaire \citep{Felder1997}, which allows for obtaining the developer's learning style\citep{Felder1988}. We only concentrate on the \emph{processing} dimension of the model because this dimension has shown to be strongly related to Scrum \citep{Scott2014, Scott2013}. After completing the questionnaire, we know that \emph{John} is an \emph{active} developer whose learning preferences are aligned to studying concepts with real examples and conduct group activities; in contrast, \emph{Mary} is a \emph{reflective} developer who prefers to receive theoretical explanations of the concepts and have space to reflect on them \citep{Felder1988}.

In the second step \ding{203}, students receive tailored instruction
according to their learning styles. The environment has a database with a predefined set of Scrum topics to tailor the instruction. That is, the environment can modify the user interface according to predefined teaching strategies suitable for active and reflexive students. For instance, given the topics on User Storie's definition, these contents are delivered John (\emph{active}) using \emph{active} strategies. They include showing a brief explanation of User Stories and their syntax, and delivering an activity that encourages John to create two User Stories from a narrative by interacting with other learners. On the other hand, Mary (\emph{reflective}) receives the same content on User Stories definition but using \emph{passive} strategies. They include showing the explanation on User Stories and their syntax and then triggering a question to reflect on the topic: "What could be the different roles, desires and benefits of the User Stories?". Moreover, the coach might give additional explanations and examples on the topic or even answer questions. 

\begin{figure}
\centering
\subfloat[Learning task tailored to John's learning style.\label{fig:Tarea-de-aprendizaje-juan}]{%
\resizebox*{7cm}{!}{\includegraphics{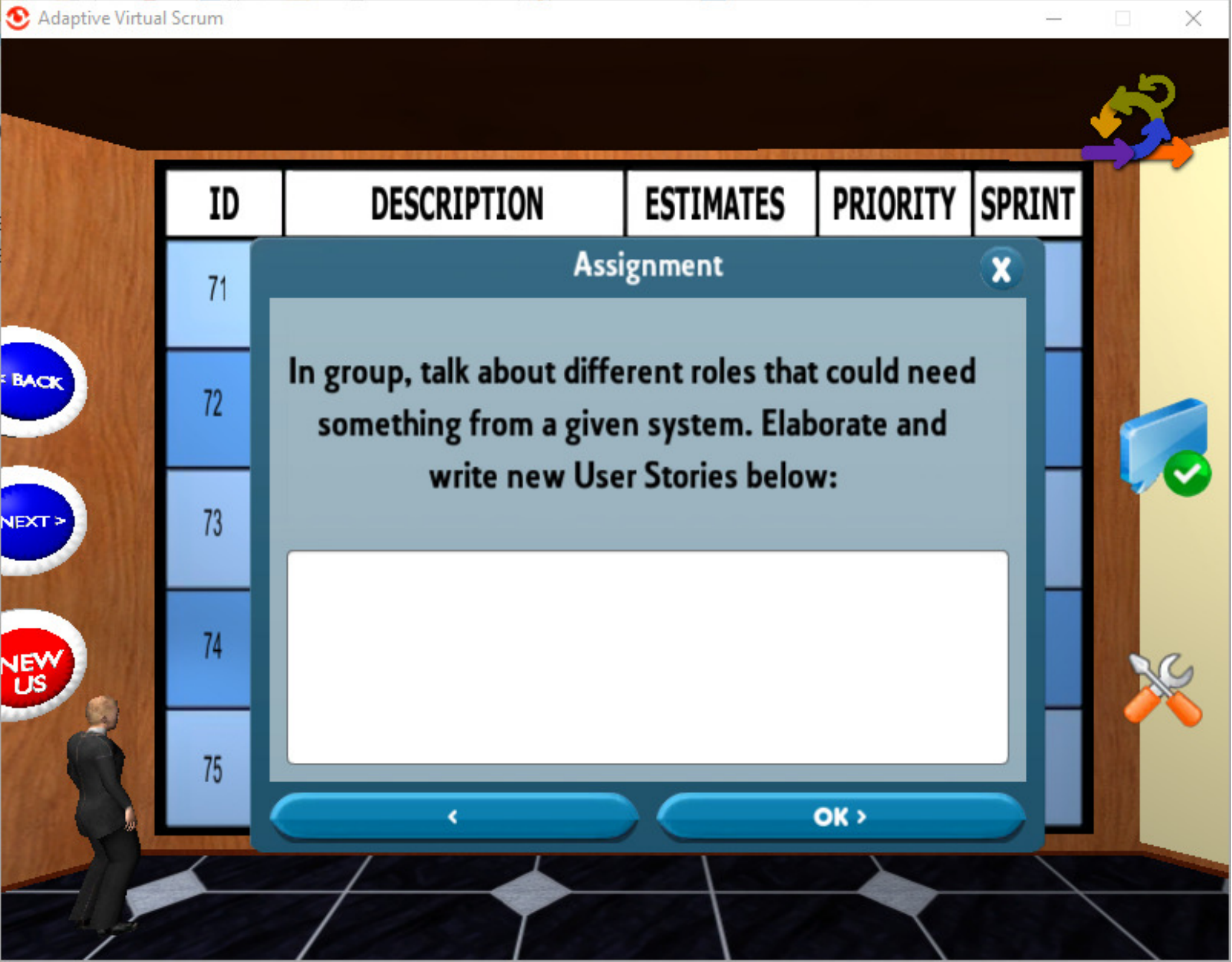}}}\hspace{5pt}
\subfloat[Learning task tailored to Mary's learning style.\label{fig:Tarea-de-aprendizaje-maria}]{%
\resizebox*{7cm}{!}{\includegraphics{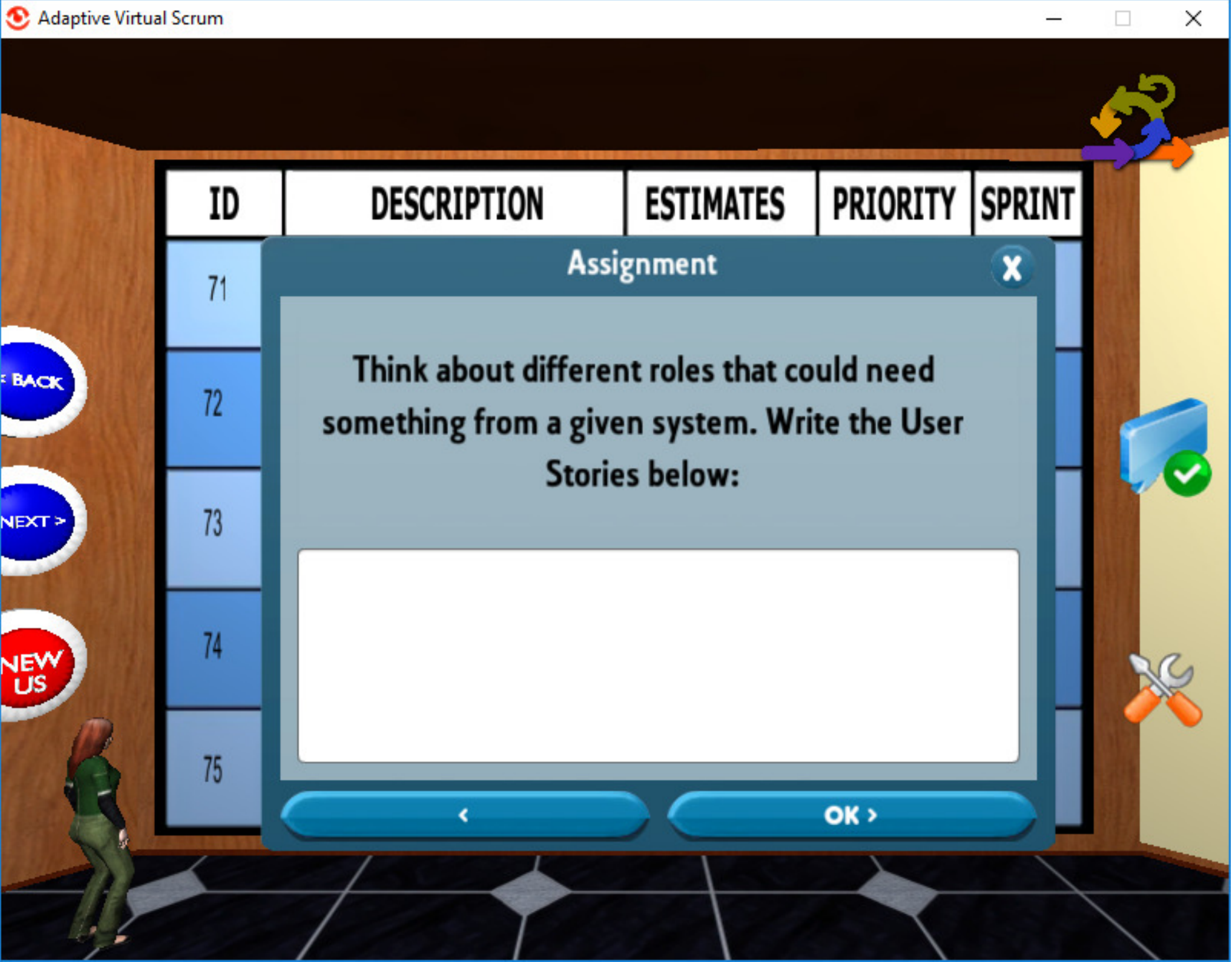}}}
\caption{Example of a learning tasks tailored to two different learning styles.} \label{fig:Prototipo-de-tareas-de-aprendizaje}
\end{figure}

Figure \ref{fig:Prototipo-de-tareas-de-aprendizaje} depicts the environment delivering contents to the developers. In particular, Figure \ref{fig:Tarea-de-aprendizaje-juan} shows the scenario in which John receive a brief explanation on User Stories and immediately is encouraged to discuss the building of User Stories with other developers. In contrast, Figure \ref{fig:Tarea-de-aprendizaje-maria} shows the scenario in which Mary receive a deeper explanation of building User Stories and immediately how she is asked a question that allows her to reflect on the concept. 

Step \ding{204} involves fostering practical experience to strengthen the developer's comprehension of Scrum. Our approach encourages developers to carry out a capstone project. The coach delivers to the developers the requirements for the capstone project in the form of Epics. This step requires that both John and Mary work in teams by following the Scrum practices learned before. 
Additionally, developers use the environment to track all the information about the execution of the practices, such as the User Stories they define, the story points they plan, the estimation values, and the status of the user stories they develop. At the end of the Sprint, teams provide an increment of the required functionality that is shown at the Sprint Review meeting. 

After the Sprint Review meeting, the coach has the opportunity to give developers valuable feedback on their performance and explain the issues that eventually arise from the development process. Moreover, the developers and the coach can analyse the traditional Scrum metrics such as work capacity, the accuracy of estimation, and burn-down charts by using the environment. Once the feedback is given, both developers John and Mary are ready to continue moving through the next Sprints by the same process. 

\section{The adaptive virtual environment}\label{sec:the-environment}

Adaptive virtual learning environments are computer-based systems that support the user in acquiring knowledge by adapting the system to the individual user \citep{Wenger2014, Kavcic2000}. To be able to adapt itself to an individual user, the system has to be aware of the individual characteristics of the user as well as the teaching domain. Moreover, the system has to have a mechanism for tailoring contents according to the user model. Hence, the user, the domain knowledge, and the adaptation mechanism can be considered essential parts of any adaptive system. In the following sections, we show how these components are represented in the context of this study.

\subsection{The learner model}

Defining the learner model involves any issues related to identifying, representing, and updating the learner's information. This model is considered to be crucial since it is the primary source used by adaptive systems to achieve the learning personalisation \citep{Brusilovsky2001, Cristea2004}. In this study, the user may be a student in a Software Engineering course or even a novice software developer in an industrial setting. Since our approach aims to align the learning process to the learner-centred principles, we consider the individual characteristics and preferences in our learner model.

To model these learner's characteristics, we rely on learning styles as useful indicators since they are defined as the characteristic cognitive, affective, and psychological behaviour that serve as relatively stable indicators of how learners perceive, interact with, and respond to the learning environment \citep{Keefe1988}. Many studies report that the usage of learning styles in teaching is a factor that can improve the quality of education \citep{Felder2005, Hawk2007}. Out of the learning style models, we use the Felder-Silverman Learning Style Model (FSLSM) \citep{Felder1988}.

Three main reasons support the selection of the FSLSM. First, previous evidence on the relationship between Scrum and learning styles in the context of an undergraduate software engineering course was found \citep{Scott2014, Scott2016}. Second, the FSLSM has been widely used in the context of Computer Science education \citep{Graf2009}. Third, the learning styles can be obtained using the Index of Learning Styles (ILS) instrument, a questionnaire based on 44 items which each student responds to according to their learning preferences \citep{Felder2005}. The instrument has shown to be reliable and valid \citep{Felder2005}, and can be administered to students by an online form.

The results obtained from the ILS distinguish the learners' preferences according to the four dimensions of the model, which are \emph{perception ($D_{1}$)}, \emph{understanding ($D_{2}$)},\emph{processing ($D_{3}$)} and \emph{input of information ($D_{4}$)}. Since the ILS instrument determines the learning styles according to a numeric scale with values ranging from -11 to +11 on each dimension, we represent the learner model using the developer's learning style. Therefore, the learner model is represented as a 4-tuple of integer values:

\begin{equation}
LS=<D_{1},D_{2},D_{3},D_{4}>\,with\,D_{i} \in Z \wedge D_{i}\in[-11,+11]
\end{equation}

In particular, our approach is based on the \emph{processing ($D_{3}$)} dimension, which is in line with the \emph{active/reflective }styles. According to the FSLSM, \emph{ active} students prefer doing tasks or talking about concepts, while \emph{reflective} students are likely to manipulate and examine the information introspectively \citep{Felder1988}.

We rely on the processing dimension because of two main reasons. First, this dimension has shown to be strongly related to the performance of Scrum practices \citep{Scott2013, scott2016towards}. Second, we aim at simplifying the learner model to achieve adaptivity straightforwardly since considering all the dimensions may hinder the definition of the instructional strategies. Notice that according to the manual for interpreting the ILS instrument \citep{Felder1997}, the \emph{active/reflective} styles are determined as:

\[
D_{3}=\begin{cases}
active & if\,D_{3}>0,\\
reflective & otherwise
\end{cases}
\]

The FSLSM presents a teaching model that defines general teaching strategies suitable for each learning style. This teaching style model classifies instructional methods according to how well the learning styles are addressed. Although these instructional methods pretend to be general and suitable for any domain of knowledge, they require teachers to tailor the concepts according to their field. For this reason, it is crucial to adequate the set of Scrum topics using the guidelines of the teaching model.

\subsection{The contents and the instructional strategies}

The contents and the instructional strategies are relevant for adaptive educational systems since they determine the learning experience achieved by the learner. Moreover, the strategies and contents and learning styles can enhance learning because the students' differences are considered when learning. In the context of this study, the teaching domain (also known as the domain model) comprises a set of topics that arise from the Scrum framework, which defines a set of software engineering practices.

The set of topics relies on an existing teaching model \citep{Scott2014a}. The authors have defined the model as the result of several years of teaching software engineering practices, including topics and instructional strategies that define the syllabus. The topics are User Story definition, User Story splitting, User Story estimation and planning, Sprint review, Daily Scrum, and Sprint retrospective meetings.  

The main goal of the syllabus is that students learn the practices suggested by the Scrum framework. In particular, the students should learn that a correct user story description must have the following elements: role, desired feature, acceptance criteria and an adequate level of granularity. Moreover, it may include non-functional requirements when necessary \citep{Cohn2004}. In case of complex user stories or epics, students should learn to apply user story splitting to obtain simpler user stories. Regarding user stories estimation, Release and Sprint plans are created by the Team during their respective planning meetings using the estimates assigned to the user stories \citep{Cohn2005, Schwaber2002}. As a rule, these estimates are described using story points, representing the complexity of a User Story. Students should learn that the Team has to estimate the number of story points that can be developed during a Sprint at the beginning of each of them.

Once a Sprint has started, the Team further decomposes each User Story into its constituent tasks that must be performed to deliver a required functionality by the end of the Sprint. Students should also understand that the team is responsible for assigning the estimated duration in hours to each task of the Sprint. At this point, students are encouraged to analyse the relationship between story points and working hours. They should become aware that this relationship is not a fixed one (e.g. 1 point does not equal 8.3 hours). Moreover, they should learn the importance of estimating tasks in hours to confirm whether they have assigned an appropriate amount of work to the Sprint. Another thing we mention is that tasks are to be carried out by one student at a time as they emerge throughout the Sprint. Regarding user stories planning, we provide the students with the well-established group estimation technique called Planning Poker, which is recommended when using agile software development methods for estimating the size of user stories and developing release and iteration plans \citep{Mahnic2012b}. 

The syllabus should be tailored according to the students' learner model to achieve adaptation. In this case, the learner model is based on the students' learning styles, in particular, the \emph{active/reflective} learning style dimension of the FSLSM. The learner model is crucial since it allows the system to select the appropriate way to deliver the content to the learner. Then, we define two instructional methods by following the guidelines of the original teaching style model \citep{Felder1988}: the passive instructional method for delivering content in a suitable way for reflective students and the active instructional method for delivering content in a suitable way for active students.

\subsection{The adaptation mechanism}

The adaptation mechanism involves the techniques used by the environment to adapt itself according to the learner model and the instructional strategies. Although the two latter ones are closely related to the resulting learning experience, the adaptation mechanism is also responsible for it. In this study, the adaptation mechanism uses the previously defined learner model based on learning styles, the contents related to the Scrum topics, and instructional strategies according to teaching styles.

\begin{figure}
\includegraphics[width=1\textwidth]{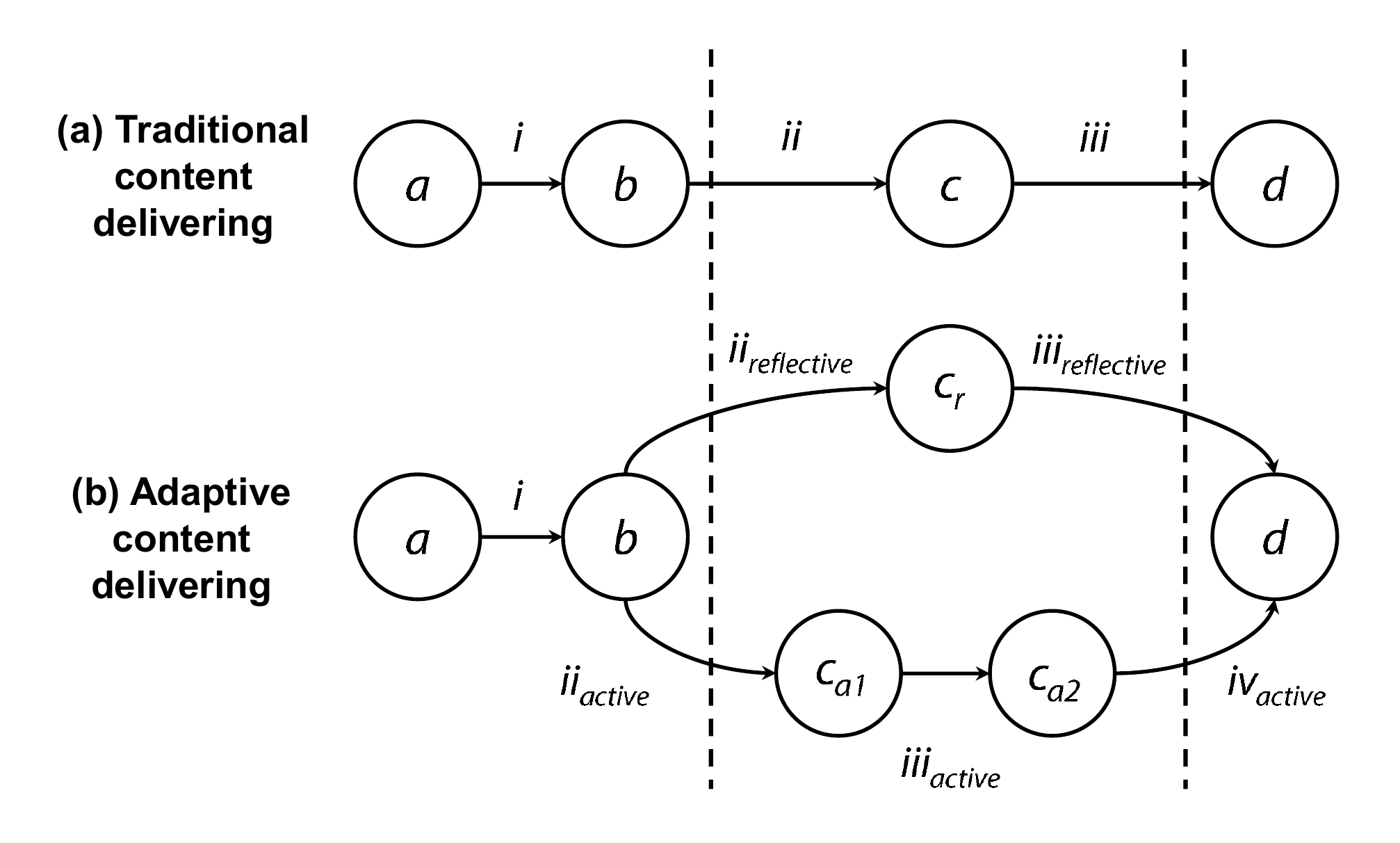}
\label{fig:Comparison-of-content-delivering}
\caption{Comparison of content delivery approaches. Nodes a, b, c, d represent the contents delivered to the learner whereas the arcs represent the way in which the contents are delivered (i.e. instructional method).}
\end{figure}

The approach requires representing the syllabus using a directed graph, where the nodes represent the topics on the syllabus, and the arcs determine the order in which topics must be delivered. Figure \ref{fig:Comparison-of-content-delivering} (a) shows an example of this structure. Since this structure represents a unique way (i.e. a unique instructional method) to deliver the topics to any student, we can say that it represents a traditional content delivery. In contrast, Figure \ref{fig:Comparison-of-content-delivering} (b) depicts an adaptive approach for delivering content, in which different ways of delivering the same topic are taken into account (i.e. two different instructional methods). As shown in the graph, some topics (nodes $a$, $b$, and $d$) could be delivered in the same way since they do not require any adaptation, whereas the topic $c$ is tailored according to the student's learning style. Thus, the original topic $c$ in the traditional delivery of content is mapped to two alternatives depending on the student's learning style: node $c_{r}$ for reflective students, and nodes $c_{a1}$ and $c_{a2}$ for active students. 

Then, for an reflective student, the topic $c$ is delivered through the active instructional method (node $c_{r}$), requiring performing the actions $ii_{reflective}$ and $iii_{reflective}$. As for an active student, the same topic $c$ is delivered through the passive instructional method (node $c_{a1}$ and node $c_{a2}$), requiring performing the actions $ii_{active}$, $iii_{active}$ and $iv_{active}$.

\begin{figure*}
\begin{centering}
\includegraphics[width=1\textwidth]{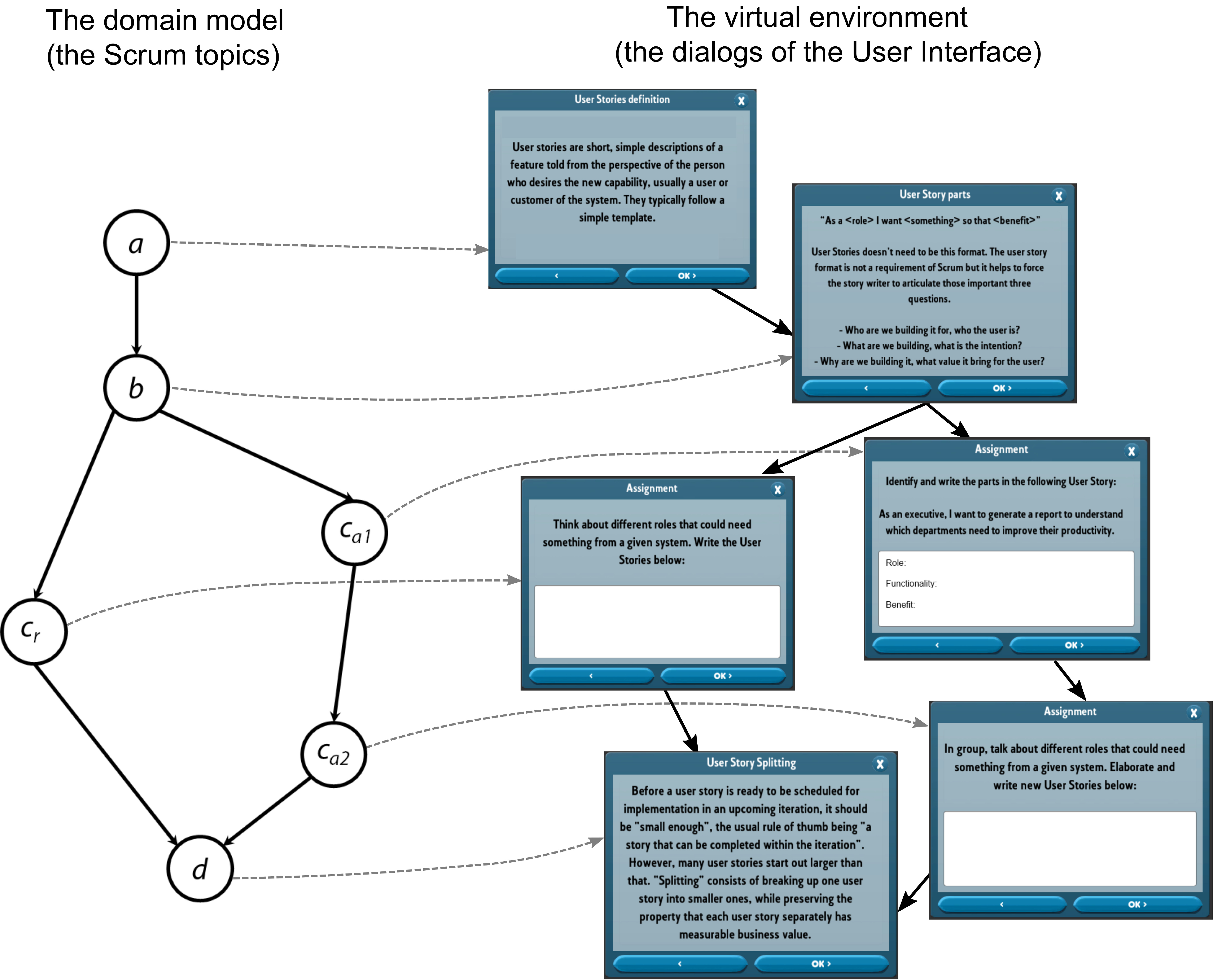}
\par\end{centering}
\caption{The key to adaptivity in the environment is the domain model behind the User Interface. \label{fig:Domain-model-and-UI}}
\end{figure*}

Let us illustrate the adaptive content delivering by the following example. For instance, the nodes $a,b,c,d$ represent a portion of the syllabus, where $a=$``User Stories definition'', $b=$``User Story parts'', $c=$``Writing User Stories'', and $d=$``User Story Splitting". Thus, the structure of the graph determines that learners should receive the topic "User Stories definition" before "User Story parts" since the latter could be difficult to understand without the former. Moreover, the topic $c=$``Writing User Stories" is taught through the passive method when the student's learning style is reflective. For this topic, this method requires three steps. First, the reflective student has to complete the previous topic (action $ii_{reflective}$). Second, the environment has to show the content (node $c_{r}=$ "Think about different roles that could need something from the system. Write the requirements as User Stories."). Third, the student has to resolve the given assignment (action $iii_{reflective}$). Figure \ref{fig:Domain-model-and-UI} shows an example of how the topics are mapped to dialogues in the User Interface of the adaptive environment.

On the other hand, the same topic $c$ is taught through the active method when the student's learning style is active. For this topic, the method requires five steps. First, the active student has to complete the previous topic (action $ii_{active}$). Second, the environment has to show the content (node $c_{a1}=$ "In group, identify and write the parts of the following User Story."). Third, the student has to complete the required information by working in a group with others (action $iii_{active}$). Fourth, the environment has to show the following content associated with topic $c$ (node $c_{a1}=$ "In group, talk about different roles that could need something from a given system. Then, elaborate new User Stories:"). Fifth, the student has to resolve the assignment by working in a group (action $iv_{active}$). In this case, the active method introduces an additional node in comparison with the passive method. 

This way, the mechanism enables the delivery of the topics in an adaptive way. More formally, we straightforwardly define the mechanism by using a rule-based approach. Although this approach is simple to implement, it has been used for most adaptive virtual learning environments \citep{Scott2016}. An adaptation rule is associated with a topic, the \emph{source topic}, and is used to control the transition to the next topic, the \emph{target topic}. Such a rule consists of two parts: a condition part and an action part. The condition part is used to specify when the learner can be directed from the current topic (the \emph{source topic}) to the next topic (the\emph{ target topic}). The condition is, in general, based on two aspects: the student's learning style and the suitability of the following topic for the learner. The learning style is stored in the learner model, whereas the suitability for the next topic is given by the underlying structure of the domain model. The action part is the consequence of the condition part, representing a predefined change in the user interface that aims at guiding the learner through the topic.

\section{Experimental evaluation}\label{sec:Evaluation}

In order to evaluate the proposed approach, we conducted a controlled experiment. The objective is to study the impact of using our adaptive virtual learning environment to support the training of developers in Scrum by taking into account their learning styles. We focused on analysing the learning gains. The data that support the findings of this study are available on request from the corresponding author. The following sections describe the study design, the measures used, and the experimental results.

\subsection{Study design}\label{sec:Experiment-2016}

The study follows the experimental design guidelines suggested by \cite{Pashler2008}, and it is organised in five main steps. First, we assign students to two groups according to their learning style: active and reflective students. We used a convenience sampling strategy to select the participants since the authors were teaching the subject of interest: Software Engineering. We used the ILS questionnaire to obtain the students' learning styles, and we only consider the processing dimension of the FSLSM. At this point, we assess the initial students' understanding of Scrum using a test (pre-test). We use an adapted version\footnote{Translated version available as supplementary material.} of the instrument initially proposed by \cite{scott2016towards}, whose quality has been analysed through item analysis.

Second, students within each learning-style group were randomly assigned to two different instructional methods: active and passive. This way, both groups have the same number of active and reflective students. This setting also allows us to compare the measures of a control group with the experimental group. Thus, the control group is given by the students who receive instruction through a method opposed to their learning style (e.g., a reflective student receives instruction based on the active method). In contrast, the experimental group is given by the set of students who receive instruction through a method that is suitable to their learning style (e.g., an active student receive instruction based on the active method) \cite{Pashler2008}.

Third, we apply Phase I of the approach. Here, the control and experimental groups receive instruction accordingly. The authors delivered the content with the support of the adaptive virtual environment. 

Fourth, after students received the training, we apply Phase II. Here, students are randomly assigned to Scrum teams (approx. five team members each), and they are asked to carry out a capstone project. The goal of the capstone course was to build a simple toy application by following the practices learnt. The teams were encouraged to use the adaptive virtual environment for managing all the artefacts needed.

Finally, after one iteration of 2 weeks, we analyse the students' understanding of Scrum topics and their performance during the capstone project. To assess the students' understanding at the end of the Sprint, we evaluate the students with the test on Scrum (post-test). Then, we calculate the students' learning gains based on the scores of the pre and post-tests.

The methods used for the data analysis were mainly descriptive statistics due to the small sample size. However, we still studied the statistical significance of the difference in the mean scores of both groups: students who were taught with suitable instructional methods (experimental group) and students who were taught in a way that did not correspond to their learning style (control group). We conducted a \textit{t-test} for independent samples \citep{Montgomery1984} and we also verified its assumptions.

\subsection{Learning gains}

We measure students' learning improvement between the beginning and end of the experience by calculating their learning gains. Equation \ref{eq:gain} shows the formula used, where the numerator gives the absolute improvement in the student's score, and the denominator is a correction factor. This factor acknowledges that it is easier for an initially low-scoring student to have a larger absolute improvement than an initially high-scoring student since they have more opportunity to change from incorrect to correct answers.

\begin{equation}
\label{eq:gain}
g_i = \frac{post_i - pre_i}{100 - pre_i}
\end{equation}

\subsection{Results}

The experiment was conducted in the first semester of 2016. Twenty-six students participated in the experiment, with 22 (85\%) identified as male and 4 (15\%) as female. Participants' ages ranged from 22 to 28 years. The participants were enrolled in the Software Engineering Workshop course, which is part of the Software Engineering curriculum at the National University of Central Buenos Aires (UNICEN). At the time of conducting the experiment, students were enrolled in university studies for 4 to 10 years, and most of the students were undergraduate (92\%). In the previous years of the curriculum, students learnt about operating systems, networking, software systems' design, object-oriented programming, and database management. Table \ref{tab:demographics} shows the demographic profile of the students.

Out of the twenty-six participants, 6 students were reflective, whereas 20 students were active. Within each style group, we divided the students into two groups according to the instructional methods. The first group contained 11 students, 9 of whom were active, and the remaining were reflective; and the second group contained 15 students, 11 of whom were active, and the remaining were reflective. Therefore, the control group (i.e., active-reflective and passive-active) and the experimental group (i.e., active-active and passive-reflective) consisted of 13 students.  

\begin{table}[]
\centering
\caption{Demographic profile of participants}
\label{tab:demographics}
\begin{tabular}{@{}llll@{}}
\toprule
Variable & Number & Percentage \\ \midrule
Gender & & \\
\qquad Female & 4 & 15\% \\ 
\qquad Male & 22 & 85\% \\ \midrule
Age & & \\ 
\qquad 22 & 4 & 15\% \\
\qquad 23 & 3 & 11\% \\
\qquad 24 & 7 & 27\% \\
\qquad 25 & 5 & 19\% \\
\qquad 26 & 5 & 19\% \\
\qquad 27 & 1 & 4\% \\
\qquad 28 & 1 & 4\% \\ \midrule
Years in university & & \\
\qquad 4 & 7 & 27\% \\
\qquad 5 & 5 & 19\% \\
\qquad 6 & 7 & 27\% \\
\qquad 7 & 5 & 19\% \\
\qquad 8 & 1 & 4\% \\
\qquad 10 & 1 & 4\% \\ \midrule
Degree & & \\
\qquad Undergraduate & 24 & 92\% \\
\qquad Postgraduate & 2 & 8\% \\ \bottomrule
\end{tabular}%
\end{table}

\begin{figure}[h]
\centering
\includegraphics[width=1\columnwidth]{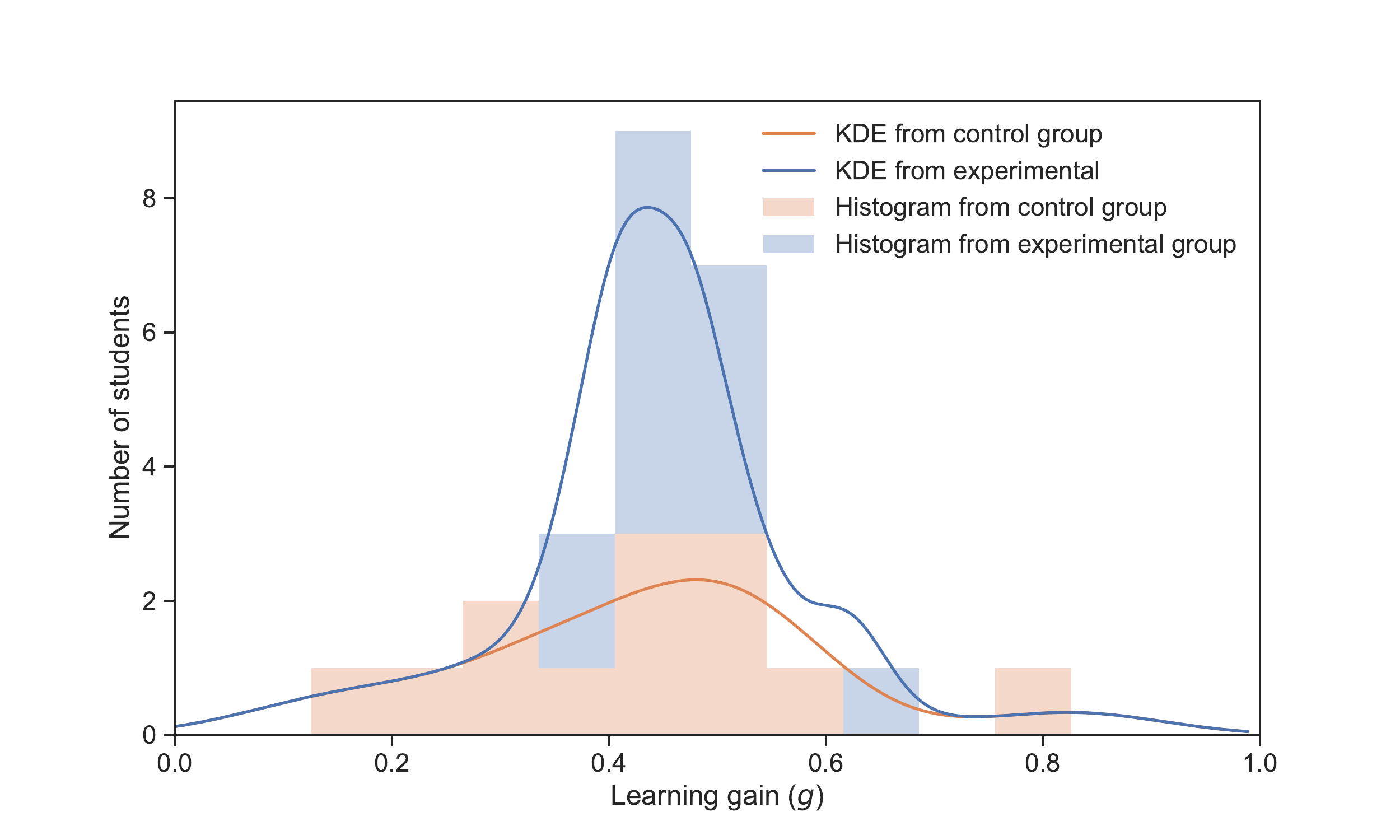}
\caption{Histograms of the learning gains of both the experimental and control groups.}
\label{fig:histograms}
\end{figure}

A comparison of the results for both the control and experimental groups is shown in Table \ref{tab:laerning-gains}. The table shows the learning gains of the groups in terms of their arithmetic mean ($M$), median ($ME$), and standard deviation ($SD$). The results show that the experimental group obtained slightly higher learning gains. Moreover, there is less variation of the learning gains in the experimental group. Figure \ref{fig:histograms} shows the histogram of the learning gains of both groups. 

Although the number of students who participate in the study represents a typical cohort size at the University of X, the sample becomes small when broken into groups. Nevertheless, we still studied the statistical significance of the results by conducting a \emph{t-test} of independent samples. We first verified its assumptions. 

The first assumption is that both samples are drawn from independent populations, which is assured by the experimental design. 
The second assumption is that the populations can be described by a normal distribution, which is verified by the shape of the density curves (KDE) shown in Figure \ref{fig:histograms} and the results from the Shapiro-Wilk normality test ($W_{Control}=0.944$, $p_{Control}=0.455$;$W_{Experimental}=0.895$, $p_{Experimental}=0.114$).

The third assumption refers to the variance homogeneity of both groups, which is supported by the results from the Levene's test ($F=4.339,$ $p=0.048$). Since there are significant differences in the variances of both groups, we conducted the \emph{t-test} considering different variances. The results of the \emph{t-test} show that there is no statistically significant difference between both means ($p=0.780$).

\begin{table}[h]
\centering
\label{tab:laerning-gains} 
\caption{Learning gains of the control and experimental groups.}
\begin{tabular}{lrrrr}
\toprule
Group       
&      M &    ME &       SD & 
n \\
\midrule
Control     
&  0.439 &  0.458 & 
0.175 &     13 \\
Experimental &  0.454 & 
0.450 &  0.067 &     13 \\
\bottomrule
\end{tabular}
\end{table}

\section{Limitations} \label{sec:Limitations}

There are several limitations in the evaluation of the proposed approach. Firstly, the notion of learning styles has been controversial for researchers \cite{Pashler2008} and should generally be treated with caution. Although there is evidence that learners have different learning preferences, there is no general agreement on the best learning style model to be used. In consequence, it might result controversial whether active and passive learning styles exist. However, the environment can be extended to support any other learning model if practitioners align with a different school of thought. 

During the definition of the learner model, we have only considered the processing dimension of the FSLSM since other studies supported the relationship between the processing dimension of the FSLSM and the way Scrum practices are performed \citep{Scott2012, scott2016towards}. However, each student has a learning style that is simultaneously described by the four FSLSM dimensions \citep{Felder1988}; then, further studies are required to analyse how the dimensions are related to the performance of Scrum.

We have also used Scrum to develop a teaching approach that covers a defined set of practices. Although these practices are general, our approach has only been applied and could only be applied to analyse students' behaviours when learning Scrum. That is, more research is needed to apply it to another agile methodology. 

Regarding the reliability and validity of the instruments used, the ILS \citep{Felder1988} has been discussed in a previous study \citep{Felder2005}. The instrument's validity used to test the students' understanding has been reviewed by the authors and one external researcher. Although the original authors of the test analysed the test quality using item analysis \citep{scott2016towards}, further studies are needed to determine the reliability of the test.

Thirdly, although we have used \textit{Virtual Scrum} as the base tool where we added our adaptive mechanism, the adaptive mechanism can be applied to other tools. For instance, project management tools such as JIRA\footnote{JIRA web site -- https://www.atlassian.com/software/jira} could be used. However, if tools were changed, data associated with the studied variables should allow for comparative analysis. Moreover, security measures should be taken into account regarding data handling and transmission. Since students/developers and project data are considered sensitive information, explicit agreement according to standards and context-dependent policies should be considered between the parties. 

Finally, it is worth noting that our approach is sensitive to the characteristics of the academic context, such as the students' motivation, the Product Owner's pressure, and the contents of the course. Another limitation is also related to the generalisation of our results; since we used a specific student sample, we can only draw conclusions for this particular population. Thus, we can state that differences in cultural and educational backgrounds could impact on the suitability of instructional activities for the students. 

\section{Conclusion\label{sec:Summary-conclusions}}

We propose an adaptive approach to deal with the problems and effort associated with personalising the training in Scrum. The central thesis of this research was that it is feasible to train software developers in Scrum by introducing an adaptive learning environment that tailors Scrum topics according to their learning style. We validated our ideas by introducing an adaptive virtual environment capable of tailoring and delivering Scrum topics to the learners according to their learning style. In particular, we used the processing dimension of the Felder-Silverman learning style model. To achieve adaptivity in the environment, we proposed an adaptation mechanism responsible for adapting the virtual environment interface taking into account the most suitable instructional method for the developer. In this sense, the proposed adaptation mechanism is the key to personalising the training process.

We conducted an experiment with Software Engineering students, and our preliminary results indicate no significant increase in the students' learning gains. However, the results show less variance in the learning gains of students who used the adaptive approach, which suggests that the approach might reduce the dispersion of the learning gains. It is worth mentioning that these results are drawn from a small sample; therefore, more research is required to show significant results when the adaptive virtual environment is used. Our approach shows great potential for providing training to software developers due to the virtual and collaborative features.

These findings provide several insights for future research. In order to generalise the findings, more experiments are needed. In this line, replicated experiments are encouraged to widen the evaluation results by studying larger samples. The application of the approach in professional contexts such as industrial companies is worth exploring. Another future research line is related to studying how variations of the approach affect students' performance. These variations could be related to using a different learner model (i.e., an alternative learning style model), different learners' performance metrics (e.g., using Scrum performance metrics), and different teaching strategies. In all cases, our approach serves as a platform to support the design of these studies. We consider that our approach represents a step towards better scrum learning using adaptive learning environments. 

\section*{Funding}
This work was supported by the National Scientific and Technical Research Council (CONICET), Argentina. Ezequiel Scott is currently founded by the Estonian Center of Excellence in ICT research (EXCITE) of the Estonian Research Council.

\bibliographystyle{apacite}
\bibliography{references.bib}

\copyrightnotice

\end{document}